\documentstyle[11pt]{article} 
\advance\hoffset by -7mm  
\renewcommand{\title}[1]{\null\vspace{0mm}

\noindent{\large{\bf #1}}\vspace{10mm}

}
\newcommand{\authors}[1]{\noindent{\large #1}\vspace{3mm}

}
\newcommand{\address}[1]{\noindent #1\vspace{5mm}

}
\renewcommand{\abstract}[1]{

\noindent
 #1}\vspace{2mm}

\begin{document}

\newcommand{\st}[2]{\stackrel{#1}{#2}}
\newcommand{\beq}{\begin{equation}}
\newcommand{\eeq}{\end{equation}}
\newcommand{\barr}{\begin{eqnarray}}
\newcommand{\earr}{\end{eqnarray}}

\newcommand{\andy}[1]{ }

%%%%%%%%%%%%%%%%%%%%%%%%%%%%%%%%%%%%%%%%%%%%%%%%%%%%%%%%%%%%%%

\title{{\bf About Synchronisation of Clocks in Free Fall Around a Central 
Body}}
\authors{\bf F. Goy}
\address{{\em Dipartimento di Fisica \\
Universit\^^ {a} di Bari  \\
Via G. Amendola, 173 \\
I-70126  Bari, Italy \\ 
E-mail:  goy@axpba1.ba.infn.it}\\
\\
June 20, 1997}
 
\abstract{The conventional nature of synchronisation is discussed in inertial 
frames, where it is found that theories using different synchronisations are 
experimentally equivalent to special relativity. In contrary, in accelerated 
systems only a theory maintaining an absolute simultaneity is consistent with 
the natural behaviour of clocks. The principle of equivalence is discussed, and 
it is found that any synchronisation can be used locally in a freely falling 
frame. Whatever the choosen synchronisation, the first derivatives of the 
metric tensor disapear and a geodesic is locally a straight line. But it is 
shown that only a synchronisation maintaining an absolute simultaneity allows 
to define time consistently on  circular orbits of a Schwarzschild metric. 
}

%\newpage
 
%-----------------------------------------------------------%

\section{Introduction}

Since a few decades there have been a revival of so-called ``relativistic 
ether theories''. This revival is partly due to the parametrised test theory of 
special relativity of Mansouri and Sexl \cite{mase:77a}
, which in contrary to the test theory 
of Robertson \cite{robe:49a}, takes explicitly the problem of synchronisation 
of distant clocks 
within an inertial frame into account. Thought its essential importance for the 
definition of time in special relativity, most modern texbooks of relativity 
treat very shortly the question of synchronisation of clocks or do not even 
mention it. The problem of synchronisation of distant clocks arose at the end 
of the 19th century from the fall of Newtonian mechanics, in which time 
was absolute and was 
defined without any reference to experiences, and in particular to procedures 
of 
synchronisation of clocks. The nature of Newtonian time, transcending any 
experimental definition was strongly critisized by Mach. On the other side, 
one had to take 
into account for the synchronisation procedure that no instantaneous action at 
distance exists in nature. In his 1905 
\cite{eins:05a}
 article 
founding the theory of relativity, Einstein influenced by the epistemological 
conceptions of Mach gave an operational definition of time: {\em ``It might 
appear possible to overcome all the difficulties attending the definition of 
``time'' by substituing ``the position of the small hand of my watch'' for 
``time''. And in fact such a definition is satisfactory when we are concerned
with defining a time exclusively for the place where the watch is located; but 
is no longer satisfactory when we have to connect in time series of events 
occuring at different places, or--what comes to the same thing-- to evaluate 
the 
times of events occurring at places remote from the watch.''} Further he wrote:
{\em ``If at the point A of space there is a clock, an observer at A can 
determine the time values of events in the immediate proximity of A by finding 
the positions of the hands which are simultaneous with these events. If there 
is at the point B of space another clock in all respects resembling the one 
at A, it 
is possible for an observer at B to determine the time values of events in the 
immediate neighbourhood of B. But it is not possible without further 
assumption to compare, in respect of time, an event at A with an event at B. 
We have so far defined only an ``A time'' and a ``B time''. We have not defined 
a common ``time'' for A and B, for the latter cannot be defined at all unless 
we 
establish {\em by definition} that the ``time'' required by light to travel 
from 
A to B equals the ``time'' it requires to travel from B to A. Let a ray of light
start at the ``A time'' $t_{A}$ from A towards B, let it at the ``B time'' 
$t_{B}$ be reflected at B in the direction of A, and arrive again at A at the 
``A time'' $t'_{A}$.

In accordance with definition the two clocks synchronize if
\andy{eq:1}
\beq
t_{B}-t_{A}=t'_{A}-t_{B}
\label{eq:1}
\eeq
We assume that this definition of synchronism is free from contradictions, and 
possible for any number of points; and that the following relations are 
universally valid:-

1. If the clock at B synchronizes with the clock at A, the clock at A 
synchronizes 
with the clock at B.

2. If the clock at A synchronizes with the clock at B and also with the clock 
at C, the clocks at B and C also synchronize with each other.''}

As Einstein underlines it himself, this is by definition that the 
time required by light to travel from A to B and from B to A is equal. 
It means that 
the one-way velocity of light is given by a convention and not by experience. 
What is known with a great precision is the (mean) two-way velocity of light, 
which obviously can be measured with only one clock and a mirror. This last is 
known with a precision of $\Delta c/c=10^{-9}$ \cite{bate:88a} and has always 
been found to be 
constant in any direction during the whole year despite of the motion of the 
earth. The one-way velocity of light, in contrary, cannot be determined 
experimentally. Let us imagine that someone would try to measure it: he 
might send a light 
ray from a clock located at A to a clock located at B, at a distance $d$ from 
A, 
and would obtain the one-way velocity of light from A to B by dividing the 
distance $d$ by the difference between the time of arrival in B and the time of 
departure from A. But in order to compute this time difference, he first needs 
clocks which are synchronised, by means of light rays whose one-way velocity
is postulated. Thus the concepts of simultaneity and one-way velocity 
of light are bound logically in a circular way.

One can of course asks himself, if other conventions which are not 
in contradiction with experiments are possible. First we rewrite equation 
(\ref{eq:1}) such 
that the ``B time'' is defined in fonction of the ``A time''. That is:
\andy{eq:2}
\beq
t_{B}=t_{A}+\frac{1}{2}(t'_{A}-t_{A})
\label{eq:2}
\eeq
Reichenbach commented \cite{reic:59a}
: {\em ``This definition is essential for the special 
theory of relativity, but is not epistemologically necessary. If we were to 
follow an arbitrary rule restricted only to the form}
\andy{eq:3}
\beq
t_{B}=t_{A}+\varepsilon(t'_{A}-t_{A})\;\;\;\;\;0<\varepsilon<1
\label{eq:3}
\eeq
{\em it would likewise be adequate and could not be called false. If the 
special theory of relativity prefers the first definition. i.e., sets 
$\varepsilon$ equal to 1/2, it does so on the ground that this definition leads 
to simpler relations.''} Among the ``conventionalists'', who agree that one can 
choose freely $\varepsilon$, are Winnie \cite{winn:70a}, Gr\"{u}nbaum 
\cite{grun:73a}, Jammer 
\cite{jamm:79a}, Mansouri and Sexl \cite{mase:77a}, Sj\"{o}din \cite{sjod:79a},
Cavalleri and Bernasconi \cite{cava:89a}, Ungar \cite{unga:91a}, 
Vetharaniam and Stedman 
\cite{vest:91a}, Anderson and Stedman \cite{anst:94a}. 
Clearly, 
different values of $\varepsilon$ correspond to different 
values of the one way-speed of light. 

A slightly different position was developed in the parametric test 
theory of special relativity of Mansouri and Sexl \cite{mase:77a}. Following 
these authors, 
we assume  that there is {\em at least one} inertial frame in which light 
behaves isotropically. We call it the priviledged frame $\Sigma$ and denote 
space and time coordinates in this frame by the letters: $(x_{0},y_{0},z_{0},
t_{0})$. 
In $\Sigma$, clocks are synchronised with Einstein's 
procedure.
We consider also an other system $S$ moving with uniform 
velocity $v<c$ along 
the $x_{0}$-axis in the positive direction. In $S$, the coordinates are written 
with 
lower case letters $(x,y,z,t)$.
Under rather general assumptions on initial and symmetry conditions on the 
two systems ($S$ and $\Sigma$ are endowed  with orthonormal axes, which 
coincide 
at time $t_{0}=0$, ... \cite{mase:77a, sell:95a}) 
the assumption that {\bf the two-way velocity of light is $c$} and furthermore
that {\bf the time 
dilation factor has its relativistic value}, one can derive
the following transformation:
\andy{eq:4}
\begin{eqnarray}
x & = & \frac{1}{\sqrt{1-\beta^{2}}}\left(x_{0}-vt_{0}\right) \nonumber \\
y & = & y_{0} \nonumber \\
z & = & z_{0} \label{eq:4} \\
t & = & s\left(x_{0}-vt_{0}\right) + \sqrt{1-\beta^{2}}\;\; t_{0}\;\;, \nonumber
\end{eqnarray}
where $\beta=v/c$. The parameter $s$, which characterizes the synchronisation 
in the $S$ frame remains unknown. Einstein's synchronisation 
in $S$ involves: 
\mbox{ $s=-v/c^{2}\sqrt{1-\beta^{2}}$} and (\ref{eq:4}) becomes a Lorentz 
boost. For a general $s$, the inverse one-way velocity of light is given
by \cite{sell:94a}:
\andy{eq:5}
\begin{equation}
\frac{1}{c_{\rightarrow}(\Theta)} = \frac{1}{c} + \left(\frac{\beta}{c} +
s\sqrt{1-\beta^{2}}\right)\cos\Theta\;\;,
\label{eq:5}
\end{equation}
where $\Theta$ is the angle between the $x$-axis and the light ray in $S$.
$c_{\rightarrow}(\Theta)$ is in general dependent on the direction. 
A simple case is $s=0$. 
This means from (\ref{eq:4}), that at $t_{0}=0$ of $\Sigma$ we set all clocks 
of $S$ 
at $t=0$ (external synchronisation), or that we synchronise the clocks by means 
of light rays with velocity $c_{\rightarrow}(\Theta)=c/1+\beta\cos\Theta$
(internal synchronisation). We obtain the transformation:
\andy{eq:6}
\begin{eqnarray}
x & = & \frac{1}{\sqrt{1-\beta^{2}}}\left(x_{0}-vt_{0}\right) \nonumber \\
y & = & y_{0} \nonumber \\
z & = & z_{0} \label{eq:6} \\
t & = & \sqrt{1-\beta^{2}}\;\; t_{0}\;\;, \nonumber
\end{eqnarray}
This transformation maintains an absolute simultaneity (the one of $\Sigma$)
between all the inertial frames.
It should be stressed that, unlike to the parameters of length contraction 
and time dilation, {\em the parameter $s$ cannot be tested}, but its 
value must be 
assigned in accordance with the synchronisation choosen in the experimental 
setup. It means, as regards experimental results, that theories using 
different $s$ are equivalent. Of course, they may predict different values of
physical quantities (for example the one-way speed of light). This 
difference resides not in nature itself but in the convention used for the 
synchronisation of clocks. With other words two transformations (\ref{eq:4}) 
with different $s$ represent the same transformation but relative to different 
time coordinates.
For a recent and comprehensive discussion of this subject, see 
\cite{vest:93a}. A striking consequence of (\ref{eq:6}) is that the negative 
result 
of the Michelson-Morley experiment does not rule out an ether. Only an ether 
with galilean transformations is excluded, because the galilean transformations 
do not lead to an invariant two-way velocity of light in a moving system.

Strictly speaking, the conventionality of clock synchronisation was 
only shown to hold in inertial frames. The derivation of equation 
(\ref{eq:4}) is done in inertial frames and is based on the assumption that 
the two-way velocity of light is constant in all directions. This last 
assumption is no longer true in accelerated systems. But special relativity is 
not only used in inertial frames. A lot of textbooks bring examples 
of calculations done in accelerated systems, using infinitesimal  Lorentz 
transformations. Such calculations use an additional assumption: the so-called 
{\em Clock Hypothesis}, which states that seen from an inertial frame, 
the rate of an accelerated ideal 
clock is identical to that of the instantaneously comoving inertial frame. With 
other words the rate of a clock is not influenced 
by acceleration per se.  
This hypothesis first used implicitely by Einstein in his article of 1905 was 
superbly confirmed in the famous timedecay experiment of muons in the CERN, 
where the muons had an acceleration of $10^{18}g$, but where their timedecay 
was only due to their velocity \cite{bail:77a}. We stress here 
the logical independence of this assumption from the structure of special 
relativity as well as from the assumptions necessary to derive (\ref{eq:4}). 
The opinion of the author is that the {\em Clock Hypothesis}, added to special 
relativity 
in order to extend it to accelerated systems leads to logical contradictions 
when the question of synchronisation is brought up. This idea was also 
expressed by Selleri \cite{sell:96a}.
The following example (see \cite{mast:93a}) shows it: imagine that 
two distant clocks are screwed on an inertial frame (say a train at rest) and 
synchronised with an 
Einstein's synchronisation. We call this rest frame $\Sigma$. The train 
accelerates 
during a certain time. After that, the acceleration stops and the train has 
again an inertial motion (sytem S). During acceleration, the clocks are 
submitted exactly 
to the same influences, so they have at all time the same rate, so they remain 
synchronous respective to $\Sigma$. Because of the relativity of simultaneity 
in special relativity, where an Einstein's procedure is applied to the 
synchronisation of clocks in all inertial frames, they are no more Einstein 
synchronous in $S$.   {\em So the 
Clock 
Hypothesis is inconsistent with the clock setting of relativity}. On the
other hand, the {\em Clock Hypothesis} is tested with a high degree of 
accuracy \cite{eisl:87a} 
and cannot be rejected, so one has to reject the clock setting of special 
relativity. The only 
theory which is consistent with the {\em Clock Hypothesis} is based on 
transformations (\ref{eq:4}) with $s=0$.

This is an ether theory. 
The fact that only an ether 
theory is consistent with accelerated motion gives strong evidences that an 
ether exist, but does not involve inevitably 
that our velocity relative to the ether is measurable. The opinion of the 
author is that it cannot be measured  because (\ref{eq:6}) represents another 
coordinatisation of the Lorentz transformation (obtained by clock 
resynchronisation). This last avoid as a principle 
any detection of an uniform motion through the ether. By changing the 
coordinate system, one cannot obtain a physics in which new physical 
phenomenons 
appear. But we can obtain a more consistent description of these phenomenons.

In all the considerations above, space-time was flat and no 
gravitational forces were present. In the following we want to treat the 
question of synchronisation of clocks in the framework of general relativity 
were special relativity is only valid locally. In section 2, we calculate the 
equations of motion for circular orbits in a Schwarzschild metric. In section 3,
we treat the problem of synchronisation of clocks on these orbits, and discuss
the compatibility of different synchronisations with the principle of 
equivalence.

\section{Circular orbits in a Schwarzschild metric}

In a system of reference $R$ with  coordinates $S$, $(x^{0},x^{1},x^{2},x^{3})=
(ct,r,\varphi,\theta)$ ($\theta$ is the  azimutal angle)
the spherical symmetric solution of Einstein's equations in vacuum, with the 
boundary 
condition that the metric becomes Minkowskian at infinity is the Schwarzschild 
metric:
\andy{eq:7}
\beq 
ds^{2}=-(1-\frac{\alpha}{r})(dx^{0})^{2}+(1-\frac{\alpha}{r})^{-1}dr^{2}
+r^{2}(\sin^{2}\theta d\varphi^{2} + d\theta^{2})\;\;\;,
\label{eq:7}
\eeq
were $\alpha=2GM/c^{2}$ is the Schwarzschild radius of the field of total energy
$Mc^{2}$ and $G$ the gravitational constant.
We will consider in the following only geodesics of test particles of mass $m$
with $r>\alpha$ so that we are not concerned here with the breakdown of the 
coordinate 
system at $r=\alpha$.
A Lagrangian function can be written as:
\andy{eq:8}
\beq
{\cal L}=-mc \sqrt{-g_{ij}\frac{dx^{i}}{d\tau}\frac{dx^{j}}{d\tau}}
\label{eq:8}
\eeq
and the Lagrange equations by
\andy{eq:9}
\beq
\frac{\partial {\cal L}}{\partial x^{i}}=
\frac{d}{d\tau}\frac{\partial {\cal 
L}}{\partial (\frac{dx^{i}}{d\tau})} 
\;\;\;i=0,\ldots,3.
\label{eq:9}
\eeq
The variables $x^{0}$, $\theta$, $\phi$ are cyclic and their conjuged momentum
are conserved. We can take without lost of generality: $\theta=\pi/2$, that is 
equatorial orbits only. The energy E and angular momentum L per unit of mass 
are conserved quantities:
\andy{eq:10}
\barr
L&=&r^{2}\frac{d\varphi}{d\tau}\nonumber \\ 
E&=& c\frac{dx^{0}}{d\tau}(1-\frac{\alpha}{r})
\label{eq:10}
\earr
From (\ref{eq:9}) and (\ref{eq:10}) the equation for the variable $r$ can be 
written
\andy{eq:11}
\beq
(\frac{dr}{d\tau})^{2}=\frac{E^{2}}{c^{2}}-(1-\frac{\alpha}{r})(
c^{2}+\frac{L^{2}}{r^{2}})=\frac{1}{c^{2}}(E^{2}-V^{2}(r))
\label{eq:11}
\eeq
were $V(r)$ is an effective potential. This effective potential has a local 
minimum, thus we have stable circular orbits. From  
(\ref{eq:10}), we 
then find for these circular orbits:
\andy{eq:12}
\barr
\frac{dr}{d\tau}&=& 0 \Rightarrow r=cst \nonumber \\
\frac{d\varphi}{d\tau}&=&\frac{L}{r^{2}} \Rightarrow \varphi(\tau)
=\varphi(\tau=0)+cst_{1}\tau \nonumber \\
\frac{dt}{d\tau}&=&\frac{E}{c^{2}(1-\frac{\alpha}{r})}\Rightarrow
\tau(t)=\tau(t=0)+cst_{2}t\;\;\;,
\label{eq:12}
\earr
where $cst_{1}=\frac{c}{r}\sqrt{\frac{\alpha}{2r-3\alpha}}$ and 
$cst_{2}=\sqrt{2-3\alpha/r}$.

\section{Two clocks in orbit}

We now consider a clock $\cal{A}$ in event-point $A$ $(x^{0}_{A},r_{A}, 
\varphi_{A})$ and make 
now all calculations in 1+2 dimensional space-time since we treat  
equatorial orbits only. On a circular orbit, its velocity is given by
$U=(c,0,\omega)/\sqrt{1-\alpha/r_{A}-r_{A}^{2}\omega^{2}/c^{2}}$. We 
have $U^{i}U^{j}g_{ij}=-c^{2}$ and $\omega=\frac{d\varphi}{dt}$ and is given by
the Kepler law $\omega^{2}=\frac{GM}{r_{A}^{3}}$ for circular orbits 
\cite{mith:73a}.  

The 
principle of equivalence ensure us that we can find a system of reference
$\st{o}{R}$, with a coordinate system $\st{o}{S}$ such that at event-point $A$,
$\st{o}{g_{ij}}(A)=\eta_{ij}$ and $\frac{\partial\st{o}{g_{ij}}}
{\partial \st{o}{x^{k}}}(A)=0$, where $\eta_{ij}=diag(-1,1,1)$. In particular, 
it is possible to choose a set of three mutually orthogonal unit vectors
$e^{i}_{(a)}$ such that $e^{i}_{(0)}=U^{i}/c$ and $e^{i}_{(1)}$ and 
$e^{i}_{(2)}$ fulfil the orthonormality conditions : 
$g_{ik}e^{i}_{(a)}e^{k}_{(b)}=\eta_{ab}$. Indices without parenthesis of 
$e^{i}_{(a)}$ are lowered  with $g_{ik}$, while indices with parenthesis 
are raised with $\eta^{ab}$.
We can choose $e_{(1)}$ radial and $e_{(2)}$ tangential to the orbit:
\andy{eq:13}
\barr
e_{(1)}&=&(0,\sqrt{1-\alpha/r_{A}},0)\nonumber\\
e_{(2)}&=&\frac{1}{\sqrt{1-\alpha/r_{A}-r_{A}^{2}\omega^{2}/c^{2}}}
(\frac{r_{A}\omega}{c\sqrt{1-\alpha/r_{A}}},0,
\frac{\sqrt{1-\alpha/r_{A}}}{r_{A}})
\label{eq:13}
\earr
        
The following transformation from coordinate system $S$ to $\st{o}{S}$, is such 
that the metric tensor in the new coordinates is minkowskian and his first 
derivatives disapear at point $A$ \cite[\S 9.6]{moll:72a}:
\andy{eq:14}
\beq
\st{o}{x}^{i}=e^{(i)}_{r}(x^{r}-x^{r}_{A})+\frac{1}{2}e^{(i)}_{r} 
\Gamma^{r}_{st}(A)(x^{s}-x^{s}_{A})(x^{t}-x^{t}_{A})\;\;\;i=0,\;1,\;2.
\label{eq:14}
\eeq
In the case of (\ref{eq:7}) the Christofell's symbols $\Gamma$ at $A$ are 
given by:
\andy{eq:15}
\barr 
\Gamma^{1}_{00}= \frac{1}{2}\frac{\alpha(1-\alpha/r_{A})}{r_{A}^{2}} &
\Gamma^{0}_{01}= \frac{1}{2}\frac{\alpha}{(1-\alpha/r_{A})r_{A}^{2}} &
\Gamma^{2}_{12}= \frac{1}{r_{A}}\nonumber \\
\Gamma^{1}_{11}=-\frac{1}{2}\frac{\alpha}{r_{A}^{2}(1-\alpha/r_{A})}&
\Gamma^{1}_{22}=-r(1-\alpha/r_{A})
\label{eq:15}
\earr
We obtain for the transformation between $S$ and $\st{o}{S}$:
\andy{eq:16}
\barr
\st{o}{x}^{0}&=&\frac{1}{\sqrt{1-\alpha/r_{A}-r_{A}^{2}\omega^{2}/c^{2}}}
\left[(1-\alpha/r_{A})(x^{0}-x_{A}^{0})-\frac{\omega r_{A}^{2}}{c}(\varphi-
\varphi_{A})\right.\nonumber \\
& +&\left. \frac{1}{2}\frac{\alpha}{r_{A}^{2}}(x^{0}-x_{A}^{0})(r-r_{A})-
\frac{\omega r_{A}}{c}(\varphi-\varphi_{A})(r-r_{A})
\right] \nonumber \\
\st{o}{x}^{1}&=&\frac{1}{\sqrt{1-\alpha/r_{A}}}(r-r_{A}) + 
\frac{1}{4}\frac{\alpha \sqrt{1-\alpha/r_{A}}}{r_{A}^{2}}
(x^{0}-x_{A}^{0})^{2} \nonumber \\
&-&\frac{1}{4}\frac{\alpha}{r^{2}_{A}(1-\alpha/r_{A})^{\frac{3}{2}}}
(r-r_{A})^{2}-\frac{1}{2}r_{A}\sqrt{1-\alpha/r_{A}}\;(\varphi-\varphi_{A})^{2}
\nonumber \\
\st{o}{x}^{2}&=&\frac{\sqrt{1-\alpha/r_{A}}}
{\sqrt{1-\alpha/r_{A}-r_{A}^{2}\omega^{2}/c^{2}}}
\left[-\frac{\omega r_{A}}{c}(x^{0}-x_{A}^{0})+ r_{A}(\varphi-
\varphi_{A})\right.\nonumber \\
&-& \left. \frac{1}{2}\frac{\omega \alpha}{c r_{A} (1-\alpha/r_{A})}
(x^{0}-x_{A}^{0})
(r-r_{A})+(r-r_{A})(\varphi-\varphi_{A}) \right] 
\label{eq:16}
\earr
This transformation looks like Lorentz transformation at first order, in 
particular, two distant events which are simultaneous in $\st{o}{S}$ are not 
simultaneous in $S$.
We now imagine that a clock $\cal{B}$ is located at $B$ 
$(x^{0}_{A}+dx^{0}, r_{A}, \varphi_{A}+d\varphi)$ 
and we want to synchronise it with $\cal{A}$ at $A$ using an Einstein's 
procedure. Since the 
metric is Minkowskian in $\st{o}{S}$, the velocity of light is $c$ in this 
(local) frame. The two clocks will be Einstein synchronised when: 
$\st{o}{x}^{0}_{A}=\st{o}{x}^{0}_{B}$=0. Using (\ref{eq:16}) we obtain that  
the infinitesimal time difference in $S$ $dx^{0}$ between these events is given 
by:
\andy{eq:17}
\beq 
dx^{0}=\frac{\omega r_{A}^{2}d\varphi}{c(1-\alpha/r_{A})}
\label{eq:17}
\eeq
We generalise this synchronisation procedure all along the circular orbit. It 
means that we synchronise $\cal{A}$ in $(r_{A},\varphi_{A})$, with $\cal{B}$ 
in 
$(r_{A},\varphi_{B}=\varphi_{A}+d\varphi)$, and then $\cal{B}$ with $\cal{C}$ 
located at 
$(r_{A},\varphi_{C}=\varphi_{B}+d\varphi)$, etc. If we do a whole round trip, 
we find a time lag $\Delta x^{0}$ given by:
\andy{eq:18}
\beq
\Delta x^{0}=\oint \frac{\omega r_{A}^{2}d\varphi}{c(1-\alpha/r_{A})}=
\frac{2\pi \omega r_{A}^{2}}{c(1-\alpha/r_{A})}
\label{eq:18}
\eeq
It means that $A$ is not synchronisable with itself, when we extend spatially 
the 
synchronisation procedure out of a local domain; this is clearly absurd. The 
problem occurs because $dx^{0}$ is not a total differential in $r$ and 
$\varphi$, thus the synchronisation procedure is path dependent. And in general 
one can say that if $\cal{A}$ synchronise with $\cal{B}$ then  in general 
$\cal{B}$ does not synchronise with $\cal{A}$. The same remark is valid for the 
transitivity of the relation ``is synchronous'' in the case of three clocks
$\cal{A},\cal{B}$ and $\cal{C}$   

According to Einstein in the 
citation quoted here above, it means that the definition of synchronism 
given by (\ref{eq:1}) which is free of contradictions in the case of inertial 
frames in flat space is no more free of contradiction when we want to define 
time globally in a curved space. One could think that this difficulty 
is insuperable and that it is not possible to :
\begin{enumerate}
\item Find a local inertial system such that the equivalence principle is 
respected
\item Defining time in this system in such a way that the extension out of a 
local domain of the synchronisation procedure is self consistent: ``is 
synchronised with'' is an equivalence relation.
\end{enumerate}
A similar problem occurs in the case of a rotating disk in flat space. It has 
been shown that only the transformation (\ref{eq:6}) allows a consistent 
definition of time 
on the rim of a rotating disk, while an Einstein's synchronisation leads 
to the impossibility of defining time without contradictions 
on the rim of this disk 
\cite{gose:97a}.

Guided by the experimental equivalence of relativistic ether theories and 
special relativity, we are looking for an other synchronisation of clocks in 
$\st{o}{R}$ such that the conditions 1 and 2 above are fullfilled.
The spatial part of transformation (\ref{eq:16}) is not changed by a 
resynchronisation of clocks, 
and we can again choose the vectors $e^{(1)}$, and $e^{(2)}$ as they can be 
read 
out from 
(\ref{eq:16}). We are looking for a transformation from coordinate system $S$
to local coordinate system $\hat{S}$ such that the time transformation is not 
depending on the space variables at first order. It means that $e^{(0)}$ is of 
the type $e^{(0)} = (y,0,0)$. In order to find $y$, we impone that the 
sychronisation only is different in $\hat{S}$ and $\st{0}{S}$. That is 
the rate of a clock at rest at the origin of $\hat{S}$ and $\st{o}{S}$ is 
the same when seen from $S$. From (\ref{eq:16}) we calculate easily that:
$\delta\st{o}{x}^{0}= \sqrt{1-\alpha/r_{A}-\omega^{2}r_{A}^{2}/c^{2}}\; \delta
x^{0}$, where $\delta \st{o}{x}^{0}$ is the coordinate time difference between 
two ticks of the clock in $\st{o}{S}$ and $\delta x^{0}$ is the same quantity 
in $S$.  We find 
that $y=\sqrt{1-\alpha/r_{A}-\omega^{2}r_{A}^{2}/c^{2}}$. 
Thus the transformation of the time coordinate from $S$ to $\hat{S}$ is now 
given by:
\andy{eq:19}
\beq
\hat{x}^{0}=\sqrt{1-\alpha/r_{A}-\omega^{2}r_{A}^{2}/c^{2}}(x^{0}-x^{0}_{A})+
\frac{1}{2}\frac{\alpha\sqrt{1-\alpha/r_{A}-\omega^{2}r_{A}^{2}/c^{2}}}
{(1-\alpha/r_{A})r_{A}^{2}} (x^{0}-x^{0}_{A})(r-r_{A})
\label{eq:19}
\eeq

1. Are we sure that $\hat{S}$ is a local inertial  system of coordinates? 
Yes.
The proof is indeed the same as it would be for $\st{o}{S}$. From (\ref{eq:14})
and using the fact that $e_{(r)}^{i}e^{(r)}_{j}=\delta^{i}_{j}$, we have:
\andy{eq:20}
\beq
e_{(r)}^{i}\hat{x}^{r}=(x^{i}-x^{i}_{A})+\frac{1}{2} 
\Gamma^{i}_{st}(A)(x^{s}-x^{s}_{A})(x^{t}-x^{t}_{A})\;\;\;i=0,\;1,\;2.
\label{eq:20}
\eeq
Differentiating two times with respect to $\hat{x}^{k}$ and $\hat{x}^{l}$ 
gives:
\andy{eq:21}
\beq
0= \frac{\partial^{2}x^{i}}{\partial \hat{x}^{l}\partial \hat{x}^{k}} + 
\Gamma^{i}_{st}(A)
\left[\frac{\partial^{2}x^{s}}{\partial \hat{x}^{l}\partial \hat{x}^{k}}
(x^{t}-x^{t}_{A})
+\frac{\partial x^{s}}{\partial \hat{x}^{k}}
\frac{\partial x^{t}}{\partial \hat{x}^{t}}\right]
\label{eq:21}
\eeq
Thus at point A:
\andy{eq:22}
\beq
0= \frac{\partial^{2}x^{i}}{\partial \hat{x}^{l}\partial \hat{x}^{k}} + 
\Gamma^{i}_{st}(A)
\left[\frac{\partial x^{s}}{\partial \hat{x}^{k}}
\frac{\partial x^{t}}{\partial \hat{x}^{t}}\right]
\label{eq:22}
\eeq
Because of the law of transformation of Christoffel's symbols, this mean that:
$\hat{\Gamma}^{i}_{kl}(A)=0$. So in $\hat{S}$ at $A$, a geodesic becomes a 
straight line:
\andy{eq:23}
\beq
\frac{d^{2}\hat{x}^{k}}{d\lambda^{2}}+\hat{\Gamma}^{k}_{il} 
\frac{d\hat{x}^{l}}{d\lambda}\frac{d\hat{x}^{i}}{d\lambda}=
\frac{d^{2}\hat{x}^{k}}{d\lambda^{2}}=0
\label{eq:23}
\eeq

2. Can time be defined consistently on the whole circular orbit? Yes.
We treat again the problem of synchronising a clock $\cal{A}$ at $A$ 
$(x^{0}_{A}, r_{A}, \varphi_{A})$ and a clock $\cal{B}$ at $B$ 
$(x^{0}_{A}+dx^{0}, r_{A}, \varphi_{A}+d\varphi)$
The two clocks are synchronised in the system of 
coordinates $\hat{S}$ if $\hat{x}_{A}^{0}=\hat{x}_{B}^{0}=0$. Then the time 
difference $dx^{0}$ between these events in $S$ calculated with (\ref{eq:19})
gives: $dx^{0}=0$. A similiar calculation as in (\ref{eq:18}) shows that 
$\Delta x^{0}=0$ for a whole round trip. Thus the time can be defined 
consistently on the orbit 
with such a synchronisation.

The metric in system $\hat{S}$ at $A$ is given by 
$e_{(a)}^{i}g_{ij}e_{(b)}^{j}=\hat{\eta}_{ab}$. We find
\andy{eq:24}
\beq
\hat{\eta}_{ab}=\left(\begin{array}{ccc}
-1 & 0 & \frac{r_{A} \omega}{c\sqrt{1-\alpha/r_{A}}} \\
0  & 1 & 0 \\ 
\frac{r_{A} \omega}{c\sqrt{1-\alpha/r_{A}}}& 0& 
1-\frac{r_{A}^{2}\omega^{2}}{c^{2}(1-\alpha/r_{A})} \\
\end{array}
\right)
\label{eq:24}
\eeq
In the case where the vector potential $\hat{\eta}_{0\alpha};\;\alpha=1,2$ is 
different from zero, the spatial part of the metric is not only given by the 
space-space coefficients of the metric but by $\hat{\gamma}_{\alpha\beta}=
\hat{\eta}_{\alpha\beta}-
\frac{\hat{\eta}_{0\alpha}\hat{\eta}_{0\beta}}{\hat{\eta}_{00}}$. In our 
case we have $\hat{\gamma}_{\alpha\beta}=\delta_{\alpha\beta}$. Thus the 
spatial system of coordinates is orthonormal. 
The velocity of light $c(\Theta)$  is found by solving the equation 
$ds^{2}=\hat{\eta}_{ab}d\hat{x}^{a}d\hat{x}^{b}=0$. We find that:
\andy{eq:25}
\beq
c(\Theta)=\frac{c}{1+\frac{r_{A}\omega \cos\Theta}{c \sqrt{1-\alpha/r_{A}}}}
\;\;\;'
\label{eq:25}
\eeq
where $\Theta$ is the angle between the light ray and the $\hat{x}^{2}$-axis

\section{Remarks}

1. The transformation of the time variable can easily be generalised to all 
synchronisations with a parameter $s$ like in (\ref{eq:4}):
\andy{eq:26}
\beq
x^{0}(s)=\sqrt{1-\alpha/r_{A}-\omega^{2}r_{A}^{2}/c^{2}}(x^{0}-x^{0}_{A})
+s\left[r_{A}(\varphi-\varphi_{A})-\frac{r_{A}\omega}{c}(x^{0}-x^{0}_{A})
\right]+O(x^{i}-x_{A}^{i})^{2}
\label{eq:26}
\eeq
The transformation (\ref{eq:19}) is given by $s=0$ and  
(\ref{eq:16}) by 
$s=-\frac{\omega r_{A}}{c\sqrt{1-\alpha/r_{A}-\omega^{2}r_{A}^{2}/c^{2}}}$
A similar argument as in section 3 shows that only $s=0$ lead to $\Delta 
x^{0}=0$ for a whole round trip of synchronisation around the orbit.

2. The inertial coordinate systems $\st{o}{S}$ and $\hat{S}$ are different 
coordinatisations of the same reference frame $\st{o}{R}$. The transformation 
from $\st{o}{S}$ to $\hat{S}$ does not involve time in the 
transformation of space variables and thus is what M\mbox{\o}ller 
\cite[p. 267, 316]{moll:72a} calls
a linear gauge transformation.

3. If a clock $\cal{A}$ at $A$ $(x^{0}_{A},r_{A},\varphi_{A})$ and a 
clock $\cal{B}$ at $B$ $(x^{0}_{A}+dx^{0},r_{A},\varphi_{A}+d\varphi)$ are 
Einstein's synchronised in the system $\st{o}{S}$ of section 3 
(i.e $dx^{0}$ is given by (\ref{eq:17})), they remain Einstein's synchronised 
during their trip around the orbit. From the equation of motion (\ref{eq:12})
one sees that they will be at a later time at point $\tilde{A}$ and $\tilde{B}$
with coordinates in $S$: $(x^{0}_{\tilde{A}},r_{\tilde{A}},\varphi_{\tilde{A}})$
and $(x^{0}_{\tilde{B}}+dx^{0},r_{\tilde{A}},\varphi_{\tilde{A}}+d\varphi)$. We 
can take a local inertial system at $\tilde{A}$ and from (\ref{eq:16}) one sees 
that : $\tilde{x}^{0}_{\tilde{A}}=\tilde{x}^{0}_{\tilde{B}}=0$.

\section{Conclusion}

In flat space, a whole set of theories equivallent to special relativity can be 
constructed. These theories are obtained by adopting an other convention on the 
synchronisation of clocks. In accelerated systems, only
the theory maintaining an absolute simultaneity is logically consistent with 
the natural behaviour of clocks.

In general relativity, the principle of equivalence tells us  that at every 
space-time point one can choose a local coordinate system such that the metric 
is minkowskian and its first derivatives disapear. Thus the laws of special 
relativity are locally valid in general relativity. In this local frame, we can 
choose an other synchronisation of clocks than the Einstein's one. The frame 
is the same but the coordinatisation is different. All these coordinatisations
are locally equivallent. The transformation between them is a linear gauge 
transformation. The spatial part of the metric is orthonormal and the derivates 
of the space time metric disapear at point in question. Thus a freely falling 
body has an uniform motion in straight line, and theses local coordinate 
systems are locally inertial.

An Einstein's synchronisation lead to a contradictory definition of time when
extended out of a local domain. It was shown in this article that in the case 
of circular orbits only a transformation maintaining an absolute simultaneity 
ables to define time globally and consistently on the orbit. An observer moving 
around a central body, who does not want to adopt a contradictory definition of 
time (when extended spatially out of his local domain) must then conlude that 
the velocity of light is not constant.

\section{Acknowledgement}
I want to thanks the Physics Departement of Bari Uni\-ver\-si\-ty for 
hospitality, 
and Prof. F. Selleri for its kind suggestions and criticisms.

\end{document}